\documentclass[12pt]{iopart}
\usepackage{iopams}

\newcommand{\de}{\partial}

\newcommand{\diag}{{\rm diag}}

\newcommand{\eqref}[1]{(\ref{#1})}
\newcommand{\CC}{\mathbb{C}}

\newcommand{\dep}[2]{\frac{\partial #1}{\partial #2}}
\newcommand{\spc}[2]{\omega^#1_{\phantom{#1}#2}}

\newcommand{\eqn}{\begin{eqnarray}}
\newcommand{\feqn}{\end{eqnarray}}
\newcommand{\be}{\begin{equation}}
\newcommand{\ee}{\end{equation}}
\newcommand{\bes}{\begin{equation*}}
\newcommand{\ees}{\end{equation*}}
\newcommand{\smat}{\left( \begin{smallmatrix}}
\newcommand{\smct}{\end{smallmatrix}\right)}

\newcommand{\ident}{I}

\begin{document}

\title{Quantum properties of the Dirac field on BTZ black hole backgrounds}

\author{Francesco Belgiorno$^1$, Sergio L Cacciatori$^{2,4}$,\\ Francesco Dalla Piazza$^{2,4}$ and Oliver F Piattella$^{3,4}$}

\address{$^1$ Dipartimento di Fisica, Universit\`a degli Studi di Milano, via Celoria 16, 20133 Milano, Italy}
\address{$^2$ Dipartimento di Fisica, Universit\`a degli Studi dell'Insubria, Via Valleggio 11, 22100 Como, Italy}
\address{$^3$ Departamento de F\'isica, Universidade Federal do Esp\'irito Santo, avenida Ferrari 514, 29075-910 Vit\'oria, Esp\'irito Santo, Brazil}
\address{$^4$ INFN, sezione di Milano, via Celoria 16, 20133 Milano, Italy}
\ead{\mailto{belgiorno@mi.infn.it}, \mailto{sergio.cacciatori@uninsubria.it}, \mailto{f.dallapiazza@uninsubria.it}, \mailto{oliver.piattella@gmail.com}}

\begin{abstract}
We consider a Dirac field on a $(1 + 2)$-dimensional uncharged BTZ black hole background. We first find out the Dirac Hamiltonian, and study its self-adjointness properties. We find that, in analogy to the Kerr-Newman-AdS Dirac Hamiltonian in $(1+3)$ dimensions, essential self-adjointness on $C_0^{\infty}(r_+,\infty)^2$ of the reduced (radial) Hamiltonian is implemented only if a suitable relation between the mass $\mu$ of the Dirac field and the cosmological radius $l$ holds true. The very presence of a boundary-like behaviour of $r=\infty$ is at the root of this problem. Also, we determine in a complete way qualitative spectral properties for the non-extremal case, for which we can infer the absence of quantum bound states for the Dirac field. Next, we investigate the possibility of a quantum loss of angular momentum for the $(1 + 2)$-dimensional uncharged BTZ black hole. Unlike the corresponding stationary four-dimensional solutions, the formal treatment of the level crossing mechanism is much simpler. We find that, even in the extremal case, no level crossing takes place. Therefore, no quantum loss of angular momentum via particle pair production is allowed.
\end{abstract}

\maketitle


\section{Introduction}\label{intro}

We investigate the behaviour of Dirac fields on BTZ \cite{Banados:1992wn} black hole backgrounds. BTZ black holes are vacuum solutions of $(1+2)$-dimensional gravity with negative cosmological constant. They become particularly important especially in relation to the AdS/CFT conjecture \cite{Maldacena:1997re}, and also with respect to the attempt to explain microscopical statistical mechanics of black holes \cite{Carlip:2005zn}. Our interest consists in studying the quantum properties of the Dirac Hamiltonian on such BTZ solutions, with the aim of determining also if quantum instabilities are allowed. This is a nontrivial task in the case of standard $(1+3)$-dimensional black holes, in particular in the case of rotating solutions. For example, spontaneous loss of charge is a relevant topic in the framework of quantum effects in the field of a black hole \cite{Gibbons:1975kk, Khriplovich:1999gm}. It belongs to that class of phenomena which are due to vacuum instability in presence of an external field, with consequent pair creation. In particular, quantum-electrodynamics effects in presence of an external electric field have been a key-topic which has been extensively discussed. Being our interest oriented toward an application to black hole physics, we limit ourselves to quote two seminal papers \cite{Heisenberg:1935qt, Schwinger:1951xk}. An effective description of the pair creation phenomenon was provided by Damour, Deruelle and Ruffini in a series of papers \cite{Damour:1975pr, Deruelle:1974zy, Deruelle:1975jc} concerning the case of Kerr-Newman black holes.
On these backgrounds the Hamilton-Jacobi equations (H-J) for a classical charged particle can be easily reduced to quadrature by means of variables separation. In particular, the radial equation describes a one dimensional motion of a particle in a given potential. The H-J equation, beyond a positive energy potential, determines a negative energy potential which at the classical level must be discarded. However, at the quantum level, negative energy states must be included, and a quantum interpretation to this couple of potentials can be given. The positive energy potential determines the allowed positive energy states, whereas the negative energy potential determines the allowed negative energy states. The usual separation of these states occurring in absence of external fields is not ensured {\it a priori}, and there can be regions where an overlap of positive and negative states for the particle is allowed, i.e. the Klein paradox takes place. In these level crossing regions, by means of tunnelling between negative and positive states, pair production of charged particles can take place with a rate determined by the transmission probability for the particle to cross the forbidden region between the two potentials, and can be computed e.g. in the WKB approximation. See for example \cite{Belgiorno:1998gj,Belgiorno:2007va, Belgiorno:2008mx, Belgiorno:2009da, Belgiorno:2009pq,belcaccianova}.

In this paper we tackle a similar but in some sense more intriguing problem: the fact that created particle pairs do not carry away charge but angular momentum. In this case, the picture is much more complicated because stationary but non-static black hole solutions are involved. In general, what happens is that it is impossible to separate the radial variable from the angular one (see e.g. \cite{Finster:2000jz,finster-axi,yamada,winklyamada,
schmid,baticschmid,hafner,Belgiorno:2008hk, Belgiorno:2008xn}). As a consequence, this makes difficult to find out an explicit form for the energy potentials, though the Hamiltonian formulation is in principle possible. A major simplification which is provided by the present model consists in the fact that, differently from the above mentioned $(1+3)$-dimensional cases, a complete separation is possible, and then a neat study of the potentials and of their level crossing is available. Even if naively a pair-creation process, signalled by level crossing, would be expected on the grounds of results obtained in the Kerr-Newman case \cite{Damour:1975pr, Deruelle:1974zy, Deruelle:1975jc}, 
we can show that no level crossing occurs on the BTZ black hole background. 

In our analysis, we take into account self-adjointness properties of the Dirac Hamiltonian, and also its qualitative spectral properties. In the non-extremal case, we are able to show that no quantum bound states exist, i.e.  no time-periodic and normalizable solution of the Dirac equation is allowed. This is in agreement with the seminal studies for the Kerr-Newman $(1+3)$-dimensional solutions carried out in \cite{Finster:2000jz,finster-axi} and dealt with different tools in e.g. \cite{yamada,winklyamada}. Compare also \cite{Belgiorno:2008hk} for the Kerr-Newman-AdS case and \cite{Belgiorno:2008xn} for the Kerr-Newman-dS one. 

The paper is structured as follows. In Sec.~\ref{Sec:BTZmetric} we briefly present the BTZ metric (without charge). In Sec.~\ref{Sec:Dirac} we write down the Dirac equation and present its Hamiltonian formulation; we calculate the energy potentials, to be compared with the classical ones coming from the classical theory. In Sec.~\ref{Sec:SADirac} and Sec.~\ref{Sec:Spectran} we consider the essential self-adjointness of the reduced Dirac operator and analyse its spectral properties. Level crossing is discussed in Sec. \ref{lev-cross}. 
Section~\ref{Sec:Conc} is devoted to discussion and conclusion. In the Appendix 
we address the Klein-Gordon equation and we find that the level crossing is absent.


\section{The BTZ metric}\label{Sec:BTZmetric}

The BTZ solution for a (1+2)-dimensional rotating and uncharged black hole has the following form \cite{Banados:1992wn}:
\begin{equation}\label{BTZmet}
\rmd s^2 = -N^2\rmd t^2 + \frac{1}{N^2}\rmd r^2 + r^2\left(N_\phi \rmd t + \rmd\phi\right)^2\;,
\end{equation}
where the lapse function $N$ and the angular shift $N_\phi$ are given by
\begin{equation}\label{BTZlapseangshift}
N^2 := - M + \frac{r^2}{l^2} + \frac{J^2}{4r^2}\;, \qquad \qquad N_\phi := -\frac{J}{2r^2}\;.
\end{equation}
Here $l$ is the curvature radius and $M$ and $J$ are two integration constants associated with the asymptotic invariance under time displacements (mass) and rotations (angular momentum). The geometry of solution~\eqref{BTZmet} is discussed in detail in \cite{Banados:1992gq}.

By solving $N^2 = 0$ one finds
\begin{equation}\label{rpm}
r_{\pm} = l\left[\frac{M}{2}\left(1 \pm \sqrt{1 - \frac{J^2}{M^2l^2}}\right)\right]^{\frac 12}\;,
\end{equation}
where $r_+$ is the horizon radius. It exists iff $M > 0$ and $J^2 \leqslant M^2l^2$, being $J^2 = M^2l^2$ the extremal case.


\section{The Dirac equation on the BTZ metric}\label{Sec:Dirac}

The Dirac equation in BTZ manifolds has been discussed in \cite{Li:2008ws}, with reference to pair-creation
by tunnelling process by a black hole horizon, in the semiclassical approximation, and also in \cite{Pitelli:2008pa},
in the case of naked singularity manifolds, by means of an asymptotic development of the geometry.
The latter strategy has been adopted also in \cite{Unver:2010uw}. To our best knowledge, a complete analysis
of the Dirac equation, its separability, and the introduction of the Dirac Hamiltonian, with a complete
determination of its spectral properties in the case of an uncharged rotating BTZ black hole solution is still
lacking, and it is provided in the sequel.\\

In order to write Dirac equation on the BTZ background \eqref{BTZmet}, we find convenient to use the tetrad formalism and we choose the following {\it dreibein}:
\begin{equation}\label{triad}
\begin{array}{ll}
e^0 = N\rmd t\;, & \rmd t = N^{-1} e^0\;, \\ \\
e^1 = N^{-1} \rmd r\;, & \rmd r =N e^1\;, \\ \\
e^2 = rN_\phi \rmd t + r\rmd\phi\;, & \rmd\phi = r^{-1}e^2 - N_\phi N^{-1}e^0\;,
\end{array}
\end{equation}
which can be short-handed as $e^a = e^a{}_\mu \rmd x^\mu$, where both Latin and Greek indices span the values $\{0,1,2\}$ and $(x^0, x^1, x^2)$ correspond to $(t,r,\phi)$. Moreover, note that Latin indices are raised (or lowered) by (1+2)-dimensional Minkowski metric $\eta_{ab}=\diag\{-1,1,1\}$, whereas Greek indices by BTZ metric say $g_{\mu\nu}$ given in \eqref{BTZmet}. 

By means of the triad \eqref{triad}, we introduce the so-called generalized Dirac matrices as follows:
\begin{equation}\label{DiracMatgen}
 \gamma_\mu = e^a{}_\mu \tilde{\gamma}_a\;, \qquad \mbox{and} \qquad \gamma^\mu = e^\mu{}_a\tilde{\gamma}^a\;,
\end{equation}
where the $\tilde{\gamma}_a$'s are the usual Dirac matrices in Minkowski space. From the structure property $\{\tilde{\gamma_a},\tilde{\gamma_b}\} = 2\eta_{ab}\ident$ we obtain $\{\gamma_\mu,\gamma_\nu\} = 2g_{\mu\nu}\ident$ and combining \eqref{DiracMatgen} and \eqref{triad} we find
\begin{equation}
\begin{array}{ll}
\gamma_0 = N\tilde{\gamma}_0 + rN_\phi\tilde{\gamma}_2\;, & \gamma^0 = -N^{-1} \tilde{\gamma}_0\;,\\ \\
\gamma_1 = N^{-1}\tilde{\gamma}_1\;, & \gamma^1=N\tilde{\gamma}_1\;,\\ \\
\gamma_2 = r\tilde{\gamma}_2\;, & \gamma^2 = N_\phi N^{-1}\tilde{\gamma}_0 + r^{-1}\tilde{\gamma}_2\;.
\end{array}
\end{equation}

In dimension three a representation of the Clifford algebra is given by the usual Pauli matrices, thus we choose $\tilde{\gamma}_0  = \rmi\sigma_3$, $\tilde{\gamma}_1  =\sigma_1$ and $\tilde{\gamma}_2 =\sigma_2$, where we multiply $\sigma_3$ by the imaginary unit in order to switch from the Euclidean to
the Lorentzian signature. The Dirac equation on a general background manifold takes on the following form:
\be\label{Diraceq}
\left[\gamma^k(\partial_k-\Gamma_k)-\mu \right]\Psi=0\;,
\ee
where $\partial_k := \partial/\partial{x^k}$, $\mu$ is the mass of the Dirac particle and
\be\label{connection}
\Gamma_k=-\frac 14\gamma^j(\partial_k\gamma_j-\gamma_l\Gamma^l_{jk})\;,
\ee
is the connection. In the BTZ metric case one finds that the non vanishing Christoffel symbols are
\begin{eqnarray}
\Gamma^0_{01} =\frac 1N\dep{N}{r}-\frac{N_\phi r^2}{2N^2}\dep{N_\phi}{r}\;, \qquad \Gamma^0_{12} =-\frac{r^2}{2N^2}\dep{N_\phi}{r}\;, \\
\Gamma^1_{00} =-N^2\left(-N\dep{N}{r}+rN_\phi^2+r^2N_\phi\dep{N_\phi}{r}\right)\;,\\ 
\Gamma^1_{02} =-\frac{N^2r}{2}\left(2N_\phi+r\dep{N_\phi}{r}\right)\;, \\
\Gamma^1_{11} =-\frac 1N \dep{N}{r} \qquad \Gamma^1_{22}=-N^2r\;, \\ 
\Gamma^2_{01}=\frac{1}{2rN^2}\left(-2rNN_\phi\dep{N}{r}+r^3N_\phi^2\dep{N_\phi}{r}+2N^2N_\phi+N^2r\dep{N_\phi}{r}\right)\;,\\
\Gamma^2_{12}=\frac{1}{2N^2r}\left(r^3N_\phi\dep{N_\phi}{r}+2N^2\right)\;,
\end{eqnarray}
so that we obtain for the connection in \eqref{connection}
\begin{eqnarray}
\fl \Gamma_0 =\frac
14(\gamma^1\gamma_0\Gamma^0_{10}+\gamma^0\gamma_1\Gamma^1_{00}+\gamma^2\gamma_1\Gamma^1_{20}+\gamma^1\gamma_2\Gamma^2_{10})\;, \\
\fl \Gamma_1=\frac
14(-\gamma^0\partial_1\gamma_0-\gamma^1\partial_1\gamma_1-\gamma^2\partial_1\gamma_2+\gamma^0\gamma_0\Gamma^0_{01} \cr
\fl \phantom{\Gamma_1=\frac 14(} +\gamma^1\gamma_1\Gamma^1_{11}+\gamma^0\gamma_2\Gamma^2_{01}+\gamma^2\gamma_2\Gamma^2_{21}+\gamma^2\gamma_0\Gamma^0_{21})\;, \\
\fl \Gamma_2=\frac 14(\gamma^0\gamma_1\Gamma^1_{02}+\gamma^2\gamma_1\Gamma^1_{22}+\gamma^1\gamma_2\Gamma^2_{12}+\gamma^1\gamma_0\Gamma^0_{12})\;.
\end{eqnarray}
Finally, Dirac equation~\eqref{Diraceq} becomes
\begin{eqnarray}\label{FinDeq}
 \left\{
-\frac
1N\tilde{\gamma}_0\partial_0+N\tilde{\gamma}_1\partial_1+\left(\frac{N_\phi}{N}\tilde{\gamma}_0+\frac
1r\tilde{\gamma}_2\right)\partial_2 \right. \cr
\left. -\frac
14\left[-2\tilde{\gamma}_1\left(\dep{N}{r}+\frac Nr\right)+\ident r\dep{N_\phi}{r} \right]
-\mu\right\}\Psi=0.
\end{eqnarray}
For reference, the same result can be rapidly achieved by means of the spin connection formalism. Using the Cartan structure equation $\rmd e^a=-\spc{a}{b}\wedge e^b$ we can obtain the non-vanishing elements of the spin connection:
\begin{eqnarray}
\spc{0}{1}=\dep{N}{r}e^0-\frac r2\dep{N_\phi}{r}e^2\;, \\
\spc{0}{2}=-\frac r2\dep{N_\phi}{r}e^1\;, \\
\spc{1}{2}=-\frac r2\dep{N_\phi}{r}e^0-\frac Nre^2\;,
\end{eqnarray}
the other elements following by symmetry. Defining $2\gamma_{ab} := \left(\tilde{\gamma}_a \tilde{\gamma}_b-\tilde{\gamma}_b \tilde{\gamma}_a\right)$, Dirac equation becomes
\be
\left[\gamma^\mu\partial_\mu+\frac 14\gamma^\mu\omega_\mu{}^{ab}\gamma_{ab}-\mu\ident\right]\psi=0\;,
\ee
or, introducing the structure constant via $\left[\tilde{\gamma}_a,\tilde{\gamma}_b\right]=-2\epsilon_{ab}^{\phantom{ab}c}\tilde{\gamma}_c$ and $\epsilon_{ab}^{\phantom{ab}c}=\eta^{cd}\epsilon_{abd}$ we can write
\be \label{d2}
\left[e^{d\mu}\left(\tilde{\gamma}_d\partial_\mu-\frac 14\tilde{\gamma}_d\omega_\mu{}^{ab}\epsilon_{ab}^{\phantom{ab}c}\tilde{\gamma}_{c}\right)-\mu\ident\right]\psi=0\;.
\ee
In order to rapidly perform the computation of the second term in round brackets, we employ the following trick (due to E.~Witten). Starting from the $2$-form $\omega_\mu^{\phantom{\mu}ab}$ we define, by Hodge duality, the 1-form $2\omega_\mu^{\phantom{\mu}c} := \omega_\mu^{\phantom{\mu}ab}\epsilon_{ab}^{\phantom{ab}c}$ and
using the relation $\tilde{\gamma}_a\tilde{\gamma}_c=\eta_{ac}-\epsilon_{ac}^{\phantom{ac}d}\tilde{\gamma}_d$ we can rewrite \eqref{d2} as follows:
\be
\left[e^{a\mu}\left(\tilde{\gamma}_a\partial_\mu-\frac 12\omega_\mu^{\phantom{\mu}c}\eta_{ac}+\frac12\omega_\mu^{\phantom{\mu}c}\epsilon_{ac}^{\phantom{ac}d}\tilde{\gamma}_{d}\right)-\mu\ident\right]\psi=0\;.
\ee
Now we define the $2$-form $\Lambda^{ac}:=e^{a\mu}\omega_\mu^{\phantom{\mu}c}$ and its Hodge dual ${}^*\Lambda^d:=\Lambda^{ac}\epsilon_{ac}^{\phantom{ac}d}$, so that we obtain:
\be
\left[e^{a\mu}\tilde{\gamma}_a\partial_\mu-\frac 12\Lambda^{ac}\eta_{ac}+\frac 12{}^*\Lambda^d\tilde{\gamma}_{d}-\mu\ident\right]\psi=0\;.
\ee
Explicitly we have that
\begin{eqnarray}
\omega^0=\frac r2\dep{N_\phi}{r}e^0+\frac{N}{r}e^2\;, \\
\omega^1=\frac r2\dep{N_\phi}{r}e^1\;,\\
\omega^2=\dep{N}{r}e^0-\frac r2\dep{N_\phi}{r}e^2\;,
\end{eqnarray}
and the form $\Lambda$, using $\Lambda^{ac}=e^{a\mu}\omega_\mu^{\phantom{\mu}c}=e^{a\mu}e^b_{\phantom{b}\mu}\spc{c}{b}=\eta^{ab}\spc{c}{b}$,
where $\spc{c}{b}$ are the components of $\omega^c=\spc{c}{b}e^b=\spc{c}{b}e^b_{\phantom{b}\mu}\rmd x^\mu$, is:
\begin{eqnarray}
\Lambda^{a0}=\frac r2\dep{N_\phi}{r}\eta^{a0}+\frac Nr\eta^{a2}\;,  &&\qquad{}^*\Lambda^0=0\;,\\
\Lambda^{a1}=\frac r2\dep{N_\phi}{r}\eta^{a1}\;, &&\qquad {}^*\Lambda^1=\frac Nr+\dep{N}{r}\;,\\
\Lambda^{a2}=\dep{N_\phi}{r}\eta^{a0}-\frac r2\dep{N_\phi}{r}\eta^{a2}\;, &&\qquad {}^*\Lambda^2=0\;,\\
\Lambda^{ac}\eta_{ac} = \frac{r}{2}\dep{N_\phi}{r}\;.
\end{eqnarray}
Finally, the expression for the Dirac equation on the BTZ background is:
\be \label{diraceqf}
\left[e^{a\mu}\tilde{\gamma}_a\partial_\mu+\frac 12\left(\frac Nr+\dep{N}{r}\right)\tilde{\gamma}_1-\left(\frac r4\dep{N_\phi}{r}+\mu\right)\ident \right]\psi=0\;,
\ee
that is the same as \eqref{FinDeq}.

The expression of the Dirac conserved current $j^\mu=\bar{\psi}\gamma^\mu\psi$ suggests that the Hilbert space where the reduced Hamiltonian is formally defined is $L^2[(r_+,\infty),rN^{-1}\rmd r]^2$, where the two dimensional function 
\be
\psi(r) := \left(\begin{array}{c}\psi_1 \\ \psi_2\end{array}\right)
\ee
 is such that
\be
\int_{r_+}^\infty \rmd r\frac rN\left(|\psi_1|^2+|\psi_2|^2\right)<\infty\;.
\ee
The map 
\begin{equation}\label{themap}
 \psi\rightarrow\psi/\sqrt{Nr}\;,
\end{equation}
can reabsorb the term proportional to $\tilde{\gamma}_1$ in the Dirac equation \eqref{diraceqf}, obtaining:
\be
\left[e^{a\mu}\tilde{\gamma}_a\partial_\mu-\left(\frac r4\dep{N_\phi}{r}+\mu\right)\ident \right]\psi=0\;.
\ee
The symmetry of the BTZ metric allows to separate variables and to obtain the following reduced Hamiltonian:
\be
H_{\rm red}=\left[
\begin{array}{cc}
-N_\phi k-\frac{Nr}{4}\dep{N_\phi}{r}-N\mu & N^2\partial_1+\frac{Nk}{r}\\ \\
-N^2\partial_1+\frac{Nk}{r} & -N_\phi k+\frac{Nr}{4}\dep{N_\phi}{r}+N\mu
\end{array}\right]\;,
\label{h-red}
\ee
where $k$ is the eigenvalue of the operator $p_\phi=-\rmi\partial_2$. After the mapping \eqref{themap}, the Hilbert space in which the reduced Hamiltonian is formally defined is $L^2[(r_+,\infty),N^{-2}\rmd r]^2$ that is isomorphic to $L^2[(r_+,\infty),rN^{-1}\rmd r]^2$. The potential part of $H_{\rm red}$ can be read off \eqref{h-red} as
\be
V=\left[
\begin{array}{cc}
-N_\phi k-\frac{Nr}{4}\dep{N_\phi}{r}-N\mu & \frac{Nk}{r}\\ \\
\frac{Nk}{r} &-N_\phi k+\frac{Nr}{4}\dep{N_\phi}{r}+N\mu
\end{array}\right]\;,
\ee
whose eigenvalues are
\be\label{eigenDirac}
\lambda_{\pm}=-N_\phi k\pm\frac Nr\sqrt{\left(\frac{J}{4r}+\mu r\right)^2+k^2}\;.
\ee
Without performing the transformation \eqref{themap}, an additive term 
\be
-\frac{N^2}{4}\left(\dep{N}{r}+\frac NR\right)^2
\ee
 would have appeared under the square root.

In the next section we address the essential self-adjointness of the Dirac Hamiltonian operator \eqref{h-red}.


\section{Essential self-adjointness of the Dirac Hamiltonian operator}\label{Sec:SADirac}

First, let us define the reduced Hamiltonian on the minimal domain
$C_0^{\infty} (r_+,\infty)^2$. We have to check the essential
self-adjointness of the operator $H_{\rm red}$ with the above domain. We exploit the fact that (\ref{h-red}) is in the form of a Dirac system \cite{weidmann}, and then we can appeal to the so-called Weyl alternative.

The Weyl alternative generalized to a system of first order ordinary differential equations (\cite{weidmann}, theorem 5.6) states that the so-called limit circle case (LCC) occurs at $r=r_+$ if for every $\lambda\in \CC$ all the solutions of
$(H_{\rm red}-\lambda) g=0$ lie in $L^2[(r_+,b),N^{-2}\rmd r]^{2}$ in a right neighbourhood $(r_+,b)$ of $r=r_+$. If at least one solution not square
integrable exists for every $\lambda \in \CC$, then no boundary condition is required and the so-called limit point case (LPC) is verified. Note that, if for a fixed $\lambda_0 \in \CC$ all the solutions of $(H_{\rm red}-\lambda_0) g=0$ and of $(H_{\rm red}-\bar{\lambda}_0) g=0$ lie in $L^2[(r_+,b),N^{-2}\rmd r]^{2}$ in a right neighbourhood $(r_+,b)$ of $r=r_+$, then this holds true for any $\lambda\in \CC$ (\cite{weidmann}, theorem 5.3). The occurrence of LCC implies the necessity to introduce boundary conditions in order to obtain a self-adjoint operator. If at least one solution not square integrable exists for every $\lambda \in \CC$, then no boundary condition is required and the so-called limit point case (LPC) is verified. The same arguments  can be applied for $r=+\infty$. The Hamiltonian operator is essentially self-adjoint if the LPC is verified both at $r=0$ and at $r=\infty$ (cf. \cite{weidmann}, theorem 5.7).

In order to study the behaviour near the horizon, it is useful to introduce the tortoise-like coordinate $y$ defined by
\be
\frac{\rmd y}{\rmd r} = -\frac{1}{N^2 (r)}\;.
\ee
Then we obtain in the non-extremal case
\be\label{tort-ne}
y(r) = \frac{1}{2(r_+^2-r_-^2)} \left[ -l^2 r_+ \ln\left( \frac{r-r_+}{r+r_+}\right)+
l^2 r_- \ln\left( \frac{r-r_-}{r+r_-}\right)\right]\;,
\ee
and in the extremal one
\be
y_e (r) = -\frac{l^2}{4 r_+} \ln\left( \frac{r-r_+}{r+r_+}\right)+ \frac{l^2}{4 } \left( \frac{1}{r+r_+}+ \frac{1}{r-r_+}\right)\;;
\ee
in both cases we have put equal to zero an integration constant, in such a way that the interval $(r_+,\infty)$ is re-mapped to $(0,\infty)$, with $y\to \infty$ as $r\to r_+$. Due to this definition, the differential part of $H_{\rm red}$ becomes equal to
\be
\left[
\begin{array}{cc}
0 & -\partial_y
\\
\partial_y & 0
\end{array}
\right]\;,
\ee
and this is enough for applying the corollary to theorem 6.8 (p. 99) of \cite{weidmann} and see that the LPC occurs on the horizon. In particular, theorem 6.8 of \cite{weidmann} states that, given a Dirac system $\tau \vec{u}=B^{-1} \left[ J \vec{u}' + P \vec{u}\right]$, with $x\in (a,b)$ and 
\begin{equation}
 J=\left(
\begin{array}{cc}
0 &
1\cr
-1
& 0
\end{array}
\right)\;,
\end{equation}
if 
\begin{equation}
B=k(x) \left(
\begin{array}{cc}
1 &
0\cr
0
& 1
\end{array}
\right)\;,
\end{equation}
with $k(x)\not \in L^1 (c,b)$ for all $c\in (a,b)$, then $\tau$ is in the LPC at $b$. As a corollary, if $b=\infty$ and $k(x)=d>0$, with $d=$const., then $\tau$ is in the LPC at $b=\infty$. This holds true both in the non-extremal case and in the extremal one.\\

As to the problem at $r=\infty$, it is useful to re-write $H_{\rm red} g = \lambda g$ as a first-order differential system and then to define the variable
\be
x : = \frac{1}{r}\;,
\ee
in such a way that the aforementioned equation amounts to a first order differential system which displays a first kind singularity at $x = 0$ \cite{hsieh, walter}: one can write
\eqn
x \partial_x g = {\mathcal M} g\;,
\feqn
where the smooth matrix ${\mathcal M}$ is regular as $x \to 0^+$ and is such that $\lim_{x\to 0^+} {\mathcal M}=:{\mathcal M}_0$ is a constant matrix with eigenvalues
\be
\epsilon_\pm = \pm \mu l\;.
\ee
One can find two linearly independent solutions $g^{(1)}(x),g^{(2)}(x)$ near $x=0$ such that
\eqn
g^{(1)} (x) = x^{\epsilon_+} h_1 (x)\;,
\feqn
and
\eqn
g^{(2)} (x) =  x^{\epsilon_-} [h_2 (x)+\ln (x)\; h_3 (x)]\;,
\feqn
where
\be
h_i (x):=\left[
\begin{array}{c} 
h_{1;i} (x)\\
h_{2;i} (x)
\end{array}
\right]
\ee
are analytic near $x=0$ for $i=1,2,3$ and $h_3(x)\neq 0$ only for $\epsilon_+-\epsilon_-= 2\mu l$ integer \cite{walter}. Assuming $\mu>0$ as physically sound, and moreover by fixing $l>0$ without loss of generality, it is then easy to conclude that the limit point case \cite{weidmann} occurs at $x=0$ only for
\eqn
\mu l \geq \frac{1}{2}\;.
\label{condt0}
\feqn
It is interesting to note that \eqref{condt0} amounts to the same essential self-adjointness condition as for the Dirac Hamiltonian in Kerr-Newman-Ads black hole backgrounds \cite{Belgiorno:2008hk}.\\
\\
{\sl Boundary conditions for $\mu l < 1/2$}.   
A physically meaningful boundary condition at $r=\infty$ for the case $\mu l < 1/2$ is the so called MIT-bag boundary condition:
\be
n_{\mu} \gamma^{\mu} \psi = \psi\;,
\ee
at the boundary. This condition means that there is no flux of Dirac particles through the boundary. By taking into account that, for any boundary $r=r_0=$ constant one can easily show that the above boundary condition becomes
\be
g_1 (r_0) = g_2 (r_0)\;,
\ee
and then we have to impose
\be
\lim_{r\to \infty} g_1 (r) = \lim_{r\to \infty} g_2 (r)\;.
\ee


\section{Spectral analysis for the Dirac case}\label{Sec:Spectran}

In order to investigate the qualitative spectral properties of the Dirac Hamiltonian $H_{\rm red}$ for $\mu l\geq 1/2$, we introduce two auxiliary selfadjoint operators $\hat h_{\rm hor}$ and $\hat h_{\infty}$:
\eqn
\fl D(\hat h_{\rm hor})=\left\{ X\in L^2_{(r_{+},r_0)},\; X
\hbox{ is locally absolutely continuous}; \right. \cr
\fl \left. B(X)=0;\;
\hat h_{\rm hor} X \in L^2_{(r_{+},r_0)}\right\},\\
\fl \hat h_{\rm hor} X = H_{\rm red} X;\\
\fl D(\hat h_{\infty})\hphantom{o}=\left\{ X\in L^2_{(r_0,\infty)},\; X
\hbox{ is locally absolutely continuous};  \right. \cr
\fl \left. B(X)=0;\;
\hat h_{\infty} X \in L^2_{(r_0,\infty)}\right\},\\
\fl \hat h_{\infty} X = H_{\rm red} X.
\feqn
$r_0$ is an arbitrary point with $r_{+}<r_0<\infty$, at which the boundary condition
\be
B(X):=
\sin (\beta) X_1 (r_0)+\cos (\beta) X_2 (r_0) =0\;,
\ee
with
\be
X(r):=\left[
\begin{array}{c}
X_1 (r)\\ 
X_2 (r)
\end{array}
\right]
\ee
and with $\beta\in [0,\pi)$ is imposed. We also have defined $L^2_{(r_{+},r_0)}:= L^2\left[(r_{+},r_0), N^{-2}\rmd r\right]^2$ and $L^2_{(r_0,\infty)}:=L^2\left[(r_0,\infty), N^{-2}\rmd r\right]^2$. We first show that $\hat h_{\infty}$ has discrete spectrum and that in the non-extremal case $\hat h_{\rm hor}$ has absolutely continuous spectrum, and then we deduce qualitative spectral properties for ${\hat H}_{red}$, which is meant as the unique self-adjoint extension of $H_{\rm red}$ defined on $C_0^{\infty} (r_+,\infty)^2$. The decomposition method \cite{weidmann} is applied. See also \cite{Belgiorno:2008xn}.

\subsection{Spectrum of $\hat h_{\infty}$}

We appeal to theorem 1 p. 102 of \cite{hs-discrete}. [Note that there is a misprint in \cite{hs-discrete} regarding the condition given at p. 102, penultimate line: $-\alpha_k (r)/p_{k2}(r)$ is indicated in place of $-\alpha_k (r)/p_{k1}(r)$].

In order to follow the definitions given therein, it is useful to work with $-\hat h_{\infty}$ (cf. (1.4), p. 101 in \cite{hs-discrete}), and then we
we introduce
\eqn
p_1(r): &=& \frac{1}{N^2} (N_{\phi} k + \mu_{\phi})\;,\\
p_2(r): &=& \frac{1}{N^2} (N_{\phi} k - \mu_{\phi})\;,\\
p(r):   &=& -\frac{k}{N r}\;,
\feqn
where we have defined
\be
\mu_{\phi}(r):= \frac{Nr}{4}\dep{N_\phi}{r}+N\mu\;.
\ee
Moreover, we have
\be
\alpha_1 (r) = \frac{1}{N^2} = \alpha_2 (r)\;,
\ee
according to the notation in \cite{hs-discrete}. We also define
\eqn
p_{11}(r): &=& +\mu \frac{1}{N}\;,\\
p_{21}(r): &=& -\mu \frac{1}{N}\;,
\feqn
and $p_{12}(r),p_{22}(r)$ via $p_1(r)=p_{11}(r)+p_{12}(r)$, $p_2(r)=p_{21}(r)+p_{22}(r)$, i.e.
\eqn
p_{12}(r) = \frac{N_\phi k}{N^2} +\frac{r}{4N}\dep{N_\phi}{r} = -p_{22}(r)\;.
\feqn
There are some further definitions which are given in agreement with the hypotheses of theorem 1 in \cite{hs-discrete}:
\eqn
-\frac{\alpha_1 (r)}{p_{11}(r)} &=& - \frac{1}{\mu N} :=r_{11}(r)\;,\\
-\frac{\alpha_2 (r)}{p_{21}(r)} &=& + \frac{1}{\mu N} :=r_{21}(r)\;,
\feqn
where $r_{11}(r),r_{21}(r)$ are long-range; moreover,
\be
Q(r):= \sqrt{-p_{11}(r) p_{21}(r)} = \mu \frac{1}{N}\;,
\ee
and
\eqn
&&-\frac{p_{12}(r)}{p_{11}(r)}= - \frac{1}{\mu} \left( \frac{N_\phi k}{N}+\frac{r}{4}\dep{N_\phi}{r}\right)=:s_{13}(r)\;,\\
&&-\frac{p_{22}(r)}{p_{21}(r)}= - \frac{1}{\mu} \left( \frac{N_\phi k}{N}+\frac{r}{4}\dep{N_\phi}{r}\right)=:s_{23}(r)\;,
\feqn
where $Q(r) s_{13}(r)$ and $Q(r) s_{23}(r)$ are clearly short-range. Let
\eqn
\Delta (r) :&=& \left( \frac{1}{p_{11}(r)} \frac{\rmd}{\rmd r} p_{11}(r) - \frac{1}{p_{21}(r)} \frac{\rmd}{\rmd r} p_{21}(r)-p(r) \right) \frac{1}{Q(r)}\cr
&=& \frac{k}{\mu r} = : \Delta_1 (r)\;,
\feqn
which is long-range. Furthermore, in our case we obtain
\be
\mu_0 (r,\lambda) = \sqrt{1-\frac{1}{\mu^2 N^2}+\frac{k^2}{\mu^2} \frac{1}{r^2}}\;,
\ee
and then one can define also
\be
E(r,\lambda):= \exp \left( \int_{r_0}^r \rmd r \mu_0 (r,\lambda) Q(r)\right)\;.
\ee
It is then easy to show that the criterion for purely discrete spectrum:
\be
\int_{r_0}^\infty \rmd r\; Q(r) = \infty\;,
\ee
is satisfied, being $Q(r) \sim  \mu l/r$ as $r\to \infty$. We also point out that condition (1.5) in \cite{hs-discrete}, which is necessary
and sufficient for the LPC to occur at $r=\infty$, leads us again to (\ref{condt0}), indeed it amounts in our case to
\be
\int_{r_0}^\infty \rmd r \frac{1}{N^2(r)} E^2 (r,0)=\infty\;,
\ee
which is possible iff (\ref{condt0}) is implemented.\\

The above result implies that no contribution to the continuous spectrum arises from near $r=\infty$, and this holds true both in the non-extremal case and in the extremal one. Moreover, we can also comment that this result holds also in the case $\mu l < 1/2$, being the LCC occurring at both the extremes \cite{weidmann}.

\subsection{Spectrum of $\hat h_{\rm hor}$ for the non-extremal case}

We show that the following result holds in the non-extremal case: $\sigma_{\rm ac} (\hat h_{\rm hor}) = {\mathbb R}$. We have to distinguish between the non-extremal case and the extremal one due to a different behaviour near $r=r_+$, and we shall discuss in the following which differences occur. Let us define
\be
\varphi_+ :=\lim_{r\to r_+} (-N_\phi k) = \frac{Jk}{2 r_+^2}\;.
\ee
Note that
\be
\lim_{r\to r_+} V=\left(
\begin{array}{cc}
\varphi_+  & 0\cr
0 & \varphi_+
\end{array}
\right)\;.
\ee
We can appeal to theorem 16.7 of \cite{weidmann}, and we find that the spectrum of $\hat h_{\rm hor}$ is absolutely continuous in ${\mathbb R}-\{\varphi_+\}$.
This can be proved as follows. Let us consider the tortoise coordinate (\ref{tort-ne}) and write the potential $V[r(y)]$ as follows:
\eqn
V[r(y)]=\left(
\begin{array}{cc}
\varphi_+  & 0\cr
0 & \varphi_+
\end{array}
\right)+P_2[r(y)]\;,
\label{p1p2}
\feqn
which implicitly defines $P_2[r(y)]$. The first term on the left of \eqref{p1p2} is of course of bounded variation; on the other hand, $|P_2[r(y)]|\in L^1 (c,\infty)$, with $c\in (0,\infty)$. (We can consider for $|\cdot|$ any norm in $\mathbb{C}^2$, e.g. the Euclidean one). Indeed,
we have to check if
\eqn\label{int-weid}
\int_c^\infty \rmd y \left[ \left(-\frac{Jk}{2 r^2 r_+^2} (r^2-r_+^2) -\mu_\phi (r)\right)^2 +
2 \frac{N^2 k^2}{r^2}\right. \cr
\left. \phantom{\int_c^\infty \rmd y}+\left(-\frac{Jk}{2 r^2 r_+^2} (r^2-r_+^2) +\mu_\phi (r)\right)^2\right]<\infty\;,
\feqn
where we have left implicit $r=r(y)$. By coming back to the coordinate $r$, and keeping into account that, as $r\to r_+$, the integration measure provides a factor $(r-r_+)^{-1}$, and the square root in (\ref{int-weid}) provides a factor $\sqrt{r-r_+}$, it is evident that the above integrability
at $r=r_+$ is ensured.\\
Then the hypotheses of theorem 16.7 in \cite{weidmann} are trivially satisfied, and one finds that the spectrum of $\hat h_{\rm hor}$ is absolutely continuous in  ${\mathbb R}-\{\varphi_+\}$.\\
We have only to exclude that $\varphi_+$ is not an eigenvalue of $\hat h_{\rm hor}$ (and of the reduced Hamiltonian). We are interested in the asymptotic behaviour as $y\to \infty$ of the solutions of the linear system
\be
H_{\rm red} X = \varphi_+ X\;,
\ee
rewritten as follows:
\be
X' =:\bar{R}[r(y)] X\;,
\label{eqradialphi}
\ee
where the prime indicates the derivative with respect to $y$, and where
\be
\bar{R}[r(y)] := \left[
\begin{array}{cc}
\frac{k N}{r} & -\varphi_+ - N_{\phi}k+\mu_\phi\\ \\
\varphi_+ + N_{\phi}k+\mu_\phi & -\frac{k N}{r}
\end{array}
\right]\;.
\ee
Cf. also \cite{yamada}. 
One easily realizes that in the non-extremal case
\eqn
\int_c^{\infty} \rmd y |\bar{R}_{ij}[r(y)]|<\infty\;, \quad \forall i,j=1,2\;.
\feqn
[Indeed, each entry vanishes as $\sqrt{r-r_+}$ as $r\to r_+$, whereas the integration measure diverges as $(r-r_+)^{-1}$ in the same limit].\\
Then according to the Levinson theorem (see e.g. \cite{eastham}, Theorem 1.3.1 p.8) one can find two linearly independent asymptotic solutions as $y\to \infty$ whose leading order is given by 
\begin{equation}
 X_I =\left(
\begin{array}{c} 1\\ 0 \end{array} \right)\;, \qquad \mbox{and} \qquad X_{II} =\left(
\begin{array}{c} 0\\ 1 \end{array} \right)\;.
\end{equation}
As a consequence no normalizable solution of the equation (\ref{eqradialphi}) can exist, and then $\varphi_+$ cannot be an eigenvalue. Note that this holds true also for the case of any self-adjoint extension of the reduced Hamiltonian which is obtained, in the case $\mu l < 1/2$, by imposing local boundary conditions at $r=\infty$.\\
In the extremal case, the analysis is made more difficult because of the worst behaviour in the limit as $r\to r_+$. Compare \cite{yamada,schmid} for the case of extremal Kerr-Newman black holes. 
We leave open this problem.

\subsection{Spectrum of $\hat H$ in the non-extremal case}

If $\mu l \geq 1/2$, let us consider the complete Hamiltonian operator $H$ defined on $C_0^\infty (r_+,\infty)^2 \times {\mathcal M}_k$, where ${\mathcal M}_k$, with $k\in {\mathbb Z}$, is the subspace spanned by the orthonormal basis $\exp(\rmi k \phi)/\sqrt{2\pi}$, and let $\hat H$ be its unique self-adjoint extension for $\mu l \geq 1/2$. For $\mu l < 1/2$, let us consider any self-adjoint extension $\hat H$ obtained by imposing local boundary
conditions to the reduced Hamiltonian at $r=\infty$. In both cases,
we have the following orthogonal decomposition:
\be
\hat H = \bigoplus_{k\in {\mathbb Z}} {\hat h}_k \times I_k\;,
\ee
where $I_k$ is the unity operator on the subspace ${\mathcal M}_k$. The operators ${\hat h}_k$ correspond, for each fixed $k$, to the self-adjoint extension of $H_{\rm red}$ (whose index $k$ has been previously left implicit in order to simplify the notation). We recall that
\be
\sigma (\hat H)=\overline{\bigcup_{k\in {\mathbb Z}} \sigma ({\hat h}_{k})}\;,
\ee
and
\eqn
\sigma_{\rm p} (\hat H)=\bigcup_{k\in {\mathbb Z}} \sigma_{\rm p} ({\hat h}_{k})\;,
\feqn
and in particular
\eqn
\sigma_{\rm ac} (\hat H)=\overline{\bigcup_{k\in {\mathbb Z}} \sigma_{\rm ac} ({\hat h}_{k})}\;.
\feqn
We have found that, in the non extremal case, $\sigma ({\hat h}_{k})={\mathbb R}=\sigma_{\rm ac} ({\hat h}_{k})$, and then we can conclude that
\eqn
\sigma (\hat H)=\sigma_{\rm ac} (\hat H) = {\mathbb R}\;,
\feqn
for the non-extremal case.\\
As a corollary of our spectral analysis, we can conclude that no quantum bound states 
exist, i.e.  no time-periodic and normalizable solution of the Dirac equation is allowed in the non-extremal 
case. Then we find that also in $(1+2)$-dimensions the same phenomenon which has been pointed out 
in $(1+3)$-dimensions, first in the Kerr-Newman case \cite{Finster:2000jz,finster-axi,yamada,winklyamada}, 
and then in the Kerr-Newman-AdS and Kerr-Newman-dS cases \cite{Belgiorno:2008hk,Belgiorno:2008xn}, occurs. 


\section{Level crossing and pair-creation}
\label{lev-cross}

In black hole physics rotating solutions are considered as unstable, both because of the presence of 
an ergoregion where the Penrose process can take place, with a reduction of the black hole 
energy (see e.g. \cite{wald,frolov}), and because of a quantum instability leading to loss of 
angular momentum through spontaneous particle emission (see e.g. \cite{Damour:1975pr,frolov}). We are 
interested in the latter topic, and, recalling our discussion in Sec. \ref{intro}, 
a signal for the presence of quantum instability can be represented by the occurrence of 
level crossing between positive energy states and negative energy ones. Although this phenomenon 
is well known in the case e.g. of rotating black holes of the Kerr-Newman family in $(1+3)$-dimensions, 
difficulties arise because of the coupling between the eigenvalue equation for the angular part  
and the eigenvalue equation for the radial `reduced Hamiltonian' (see e.g. \cite{Finster:2000jz,finster-axi,yamada,winklyamada,schmid,baticschmid,hafner,Belgiorno:2008hk, Belgiorno:2008xn}). 
A relevant point of our present analysis consists just in 
checking if level crossing occurs in the present situation, where a much more simple case is at hand, 
providing us a framework which is completely under control. Unfortunately, the answer we find is 
negative, no level crossing occurs for BTZ solutions we are considering.\\

We will now show that level crossing is absent. We limit ourselves to taking into account 
the case $\mu l\geq 1/2$, because in the other case boundary conditions are expected 
to affect physical properties and require a different analysis. 
Still, a comment is in order. We note that the
expressions (\ref{eigenDirac}) for the positive and negative energy bounds are not the same as obtained in the classical limit from the Hamilton-Jacobi equations (see the Appendix).  
This is due to the presence of a term proportional to the angular momentum $J$ in the square root.
Since Dirac equation describes spin $1/2$ particles, it is not surprising to obtain such a term in the eigenvalues of \eqref{eigenDirac}. Its presence takes into account the coupling between the black hole angular momentum $J$ and the particle spin $\hbar/2$. Indeed, restoring the $\hbar$ we obtain from \eqref{eigenDirac}:
\be
\lambda_{\pm}=-N_\phi k\pm\frac Nr\sqrt{\left(\mu -N_\phi\frac{\hbar}{2}\right)^2r^2 + k^2}\;,
\ee
so that we have the expected classical $\hbar\to 0$ limit.\\
Now, as 
\be
\lambda_+(r_+)=\lambda_-(r_+)=\frac {kJ}{2r_+^2},
\ee
we will show that $\lambda_+(r)>\lambda_+(r_+)$ and $\lambda_-(r)<\lambda_+(r_+)$ for $r>r_+$. Indeed:
\begin{eqnarray}
\fl \lambda_+(r)=\frac {kJ}{2r^2} +\frac {(r^2-r_+^2)^{\frac 12}(r^2-r_-^2)^{\frac 12}}{lr^2}\sqrt {\left(\frac J{4r}+\mu r \right)^2+k^2} \cr
\fl \geq \frac {kJ}{2r^2} +|k|\frac {(r^2-r_+^2)^{\frac 12}(r^2-r_-^2)^{\frac 12}}{lr^2}
\geq \frac {|k|}{l} -\frac {|k|}{2r^2} \left( -{\rm sign}(k) J +\frac {2r_+^2}l \right)=: G_+(r),\label{diseq}
\end{eqnarray}
where we used $r_+^2\geq r_-^2$.
Now, $|J|\leq Ml$, so that using (\ref{rpm}) we see that 
\be
\frac {|J|}{2r_+^2} \leq \frac 1l. 
\ee
Thus, the parenthesis in (\ref{diseq}) define a non-negative constant which is positive 
unless the extremal case with ${\rm sign}(k){\rm sign}(J)=1$ occurs; then, when it is positive, 
$G_+(r)$ increases monotonically, so that $G_+(r_+)=\lambda_+(r_+)< G_+ (r)$ for $r>r_+$. As a consequence, 
$\lambda_+(r)> \lambda_+(r_+)$ for $r>r_+$ holds true. In the aforementioned extremal case 
with ${\rm sign}(k){\rm sign}(J)=1$ it is easy to show that $\lambda_+(r)= \lambda_+(r_+)$ 
for $r>r_+$ could occur only at the point $r_\ast := \sqrt{-J/(4\mu)}$, but the condition 
$r_\ast >r_+$ actually requires that $\mu l< 1/2$, which is not allowed under our 
assumption $\mu l\geq 1/2$. Then, we can safely conclude that a strict inequality occurs 
when no boundary condition need to be imposed.\\
In the same way we have
\begin{eqnarray}
\fl \lambda_-(r)=\frac {kJ}{2r^2} -\frac {(r^2-r_+^2)^{\frac 12}(r^2-r_-^2)^{\frac 12}}{lr^2}\sqrt {\left(\frac J{4r}+\mu r \right)^2+k^2} \cr
\fl \leq \frac {kJ}{2r^2} -|k|\frac {(r^2-r_+^2)^{\frac 12}(r^2-r_-^2)^{\frac 12}}{lr^2}
 \leq -\frac {|k|}{l} +\frac {|k|}{2r^2} \left( {\rm sign}(k) J +\frac {2r_+^2}l \right)=: G_-(r).\label{-diseq}
\end{eqnarray}
The same argument as before shows that then $\lambda_-(r)< \lambda_-(r_+)$ for $r>r_+$.\\
We then conclude there is no level crossing. We point out again that the proof works exactly in the same way for the extremal case.

\section{Summary and Conclusion}\label{Sec:Conc}

We have considered quantum properties of a Dirac field on a rotating BTZ black hole background.
By means of variable separation, we have obtained the Dirac Hamiltonian and determined that the reduced (radial)
Hamiltonian is essentially self-adjoint on $C_0^{\infty}(r_+,\infty)^2$ iff $\mu l \geq 1/2$. It is
remarkable that this condition coincides with the one for the essential self-adjointness of the reduced
Dirac Hamiltonian in the case of Kerr-Newman-AdS black hole
background. 
For the case $\mu l< 1/2$, where the reduced Hamiltonian is
not essentially self-adjoint, a particular physically
meaningful local boundary condition at $r=\infty$ is the MIT-bag boundary
condition. 
Then, we have determined spectral
properties of the Hamiltonian, and we have been able to show that its spectrum is absolutely continuous
and coincides with ${\mathbb{R}}$ in the non-extremal case. The extremal case is more difficult,
and we leave it open. In the non-extremal case, as a consequence of our analysis, we can infer that 
the point spectrum is empty, i.e. no quantum bound states 
exist.  This amounts to the absence of time-periodic and normalizable solutions of the Dirac equation. 
This matches known results for the Dirac equation on stationary  $(1+3)$-dimensional 
black hole solutions \cite{Finster:2000jz,finster-axi,yamada,winklyamada,Belgiorno:2008hk,Belgiorno:2008xn}. 
Then, in the case $\mu l \geq 1/2$, we have taken into account
the possibility of a quantum loss of angular momentum for a BTZ black hole via pair production.
To this purpose we have extracted the form of the energy potentials.
These potentials are, in the classical limit, identical to the classical calculation coming from the H-J equation.
Surprisingly, even in the extremal case no level crossing takes place so that we have not any signal that
the BTZ black hole loses angular momentum via particle pair production. 

\ack{OFP was supported by the CNPq (Brazil) contract 150143/2010-9.}

\appendix

\section{The Klein-Gordon equation on BTZ and its Hamiltonian formulation}\label{Sec:KGeqBTZ}

Consider a minimally coupled neutral scalar field $\psi(x)$. Its equation of motion is $(\square + \mu^2)\psi(x) = 0$, where $\mu^2$ is the square mass. The general form for the D'Alembertian operator on a given metric is:
\begin{equation}\label{squareop}
\square = -\frac{1}{\sqrt{-g}}\frac{\de}{\de x^{\mu}}\left(\sqrt{-g}\;g^{\mu\nu}\frac{\de}{\de x^{\nu}}\right)\;,
\end{equation}
where $g$ is the metric determinant. For the BTZ metric $g = -r^2$ and, for reference, the contravariant form of the metric in the reference frame of \eqref{BTZmet} has the following form:
\begin{equation}\label{contrmet}
g^{\mu\nu} = \left(
\begin{array}{lll}
-N^{-2} & 0 & N_\phi N^{-2}\\
0 & N^2 & 0\\
N_\phi N^{-2} & 0 & r^{-2} - N_\phi^2 N^{-2}
\end{array}
\right)\;.
\end{equation}
Computing \eqref{squareop} for the BTZ metric, the Klein-Gordon equation takes on the following form:
\begin{equation}\label{KGeqBTZ}
\fl \left[\frac{\de^2}{\de t^2} - 2N_\phi\frac{\de^2}{\de t\de \phi} - \frac{N^2}{r}\frac{\de}{\de r}\left(N^2 r\frac{\de}{\de r}\right) + \left(N_\phi^2 - \frac{N^2}{r^2}\right)\frac{\de^2}{\de\phi^2} + N^2\mu^2\right]\psi(x) = 0\;.
\end{equation}
Following, for example, \cite{Belgiorno:2008mx} we look for the following Hamiltonian formulation of \eqref{KGeqBTZ}:
\begin{equation}\label{kgveceq}
\rmi\frac{\de}{\de t}\vec{g} = H\vec{g}\;,
\end{equation}
where
\begin{equation}
 H := \left[
\begin{array}{ll}
\mathcal{V} & I\\
\mathcal{H} & \mathcal{V}^\dagger\\
\end{array}
\right]\;, \qquad \mbox{and} \qquad \vec{g} := \left(
\begin{array}{l}
 u \\
 v
\end{array}
\right)\;.
\end{equation}
The explicit form of \eqref{kgveceq} for $u$ is:
\begin{equation}\label{kgeq2}
\frac{\de^2u}{\de t^2} + \mathcal{H}u + \rmi\mathcal{V}\frac{\de u}{\de t} + \rmi\mathcal{V}^\dagger\frac{\de u}{\de t} - \mathcal{V}^\dagger\mathcal{V}u = 0\;,
\end{equation}
with the same equation also satisfied by $v$.

Comparing \eqref{KGeqBTZ} with \eqref{kgeq2} we immediately find that
\begin{equation}\label{iVVdag}
\rmi\left(\mathcal{V} + \mathcal{V}^\dagger\right) = -2N_\phi\frac{\de}{\de\phi}\;,
\end{equation}
so that $\mathcal{V}$ has the following form:
\begin{equation}\label{poteV}
\mathcal{V} = -N_\phi p_\phi = \frac{J}{2r^2}p_\phi\;,
\end{equation}
where we have introduced the momentum operator $p_\phi := -\rmi\de/\de\phi$.

From \eqref{KGeqBTZ} together with \eqref{kgeq2} and \eqref{iVVdag} we obtain that $\mathcal{H}$ has the following form:
\begin{equation}\label{hamiltonian}
\mathcal{H} = -\frac{N^2}{r}\frac{\de}{\de r}\left(N^2 r\frac{\de}{\de r}\right) -\frac{N^2}{r^2}\frac{\de^2}{\de\phi^2} + N^2\mu^2\;.
\end{equation}
After variable separation and denoting with $k$ the eigenvalue of $p_\phi$, we define the potential part of $H$ as follows (cf. also \cite{belcaccianova}):
\begin{equation}
 V := \left[
\begin{array}{cc}
\frac{Jk}{2r^2} & \ident\\ \\
N^2\left(\mu^2 + \frac{k^2}{r^2}\right) & \frac{Jk}{2r^2}\\
\end{array}
\right]\;,
\end{equation}
where $\ident$ is the identity operator. The eigenvalues of $V$ are thus
\begin{equation}\label{eigenV}
\lambda_\pm = \frac{Jk}{2r^2} \pm \frac{N}{r}\sqrt{\mu^2r^2 + k^2}\;,
\end{equation}
which represent the energy potentials with which we are about to investigate, in the next section, if the level crossing does take place. First, we show how the classical potentials deriving from the H-J equation are exactly the same as the ones in \eqref{eigenV}. We do not deal with the problem of defining an Hamiltonian 
operator in a suitable space and studying its self-adjointness properties herein. 

\section{The classical approach and the level crossing}\label{Sec:classappr}

Consider the Hamilton-Jacobi equation
\begin{equation}
g^{\mu\nu}\;\de_\mu S\;\de_\nu S + \mu^2 = 0\;,
\end{equation}
together with the variable separation $S = -\omega t + k\phi + R(r)$. One finds
\begin{equation}\label{HJseparated}
-\frac{1}{N^2}\omega^2 - 2\omega k\frac{N_\phi}{N^2} + N^2(\dot{R})^2 + \left(\frac{1}{r^2} - \frac{N_\phi^2}{N^2}\right)k^2 + \mu^2 = 0\;,
\end{equation}
where the dot denotes the derivation with respect to $r$. Recast \eqref{HJseparated} as follows:
\begin{equation}\label{deltareq}
N^4(\dot{R})^2 = \left(\omega + N_\phi k\right)^2 - N^2\left(\mu^2 + \frac{k^2}{r^2}\right)\;.
\end{equation}
It is remarkable that in \eqref{deltareq} the variable are completely separated so that we can simply solve for $\omega$ and find the following expression for the effective potentials:
\begin{equation}\label{eigenVclass}
\omega_\pm = -N_\phi k \pm \frac{N}{r}\sqrt{\mu^2 r^2 + k^2} = \frac{Jk}{2r^2} \pm \frac{N}{r}\sqrt{\mu^2r^2 + k^2}\;,
\end{equation}
which are identical to the ones in \eqref{eigenV}, as expected.\\

We show now that no level crossing occurs. However, we point out that 
what follows is reliable only if no boundary conditions have to be imposed in order to 
obtain a well-defined Hamiltonian in the quantum case.\\ 
As for the Dirac case we can show that $\lambda_+(r)>\lambda_+(r_+)$
and $\lambda_-(r)<\lambda_+(r_+)$ for $r>r_+$ and $\mu\neq 0$. Indeed, as in
section \ref{lev-cross}, with $\mu^2 r^2$ in place of $(J/4r + \mu
r)^2$, we have
\begin{eqnarray}
 \lambda_+(r)\geq \frac {|k|}{l} -\frac {|k|}{2r^2}
\left( -{\rm sign}(k) J +\frac {2r_+^2}l \right)=: G_+(r)\;,\\
\lambda_-(r)\leq -\frac {|k|}{l} +\frac {|k|}{2r^2} \left( {\rm
    sign}(k) J +\frac {2r_+^2}l \right)=: G_-(r)\;,
\end{eqnarray}
and the same arguments of section \ref{lev-cross}  lead us to the conclusion.

\end{document}